# A Strategy to enable Prefix of Multicast VoD through dynamic buffer allocation

Dr.T.R.GopalaKrishnan nair[1] and P Jayarekha [2]

[1] Director Research Industry and Incubation Centre
DSI ,Bangalore, India

[2] Dept Of ISE BMS College of Engineering,
Research Scholar Dr.MGR university,

**Abstract**

In this paper we have proposed a dynamic buffer allocation algorithm for the prefix, based on the popularity of the videos. More cache blocks are allocated for most popular videos and a few cache blocks are allocated for less popular videos. Buffer utilization is also maximized irrespective of the load on the Video-on-Demand system. Overload can lead the server getting slowed down. By storing the first few seconds of popular video clips, a multimedia local server can shield the users from the delay, throughput, and loss properties of the path between the local server and the central server. The key idea of controlled multicast is used to allow clients to share a segment of a video stream even when the requests arrive at different times. This dynamic buffer allocation algorithm is simulated and its performance is evaluated based on the buffer utilization by multimedia servers and average buffer allocation for the most popular videos. Our simulation results shows efficient utilization of network bandwidth and reduced hard disk utilization hence resulting in increase in the number of requests being served.

***Keywords:*** *Multicast transmission, Interval caching, prefix caching, Dynamic buffer Allocation.*

## 1. Introduction

Recent advances in high speed networks and communication technologies have made it possible to provide an on-line access to a variety of information sources such as reference books, journals, newspapers images and video clips. The two architectures available for servicing the client requests are Client-Pull and Server-Push. The Client-Pull type server, streams the data to the client in response to the client’s explicit request. Here the client has to determine the playback time, estimate the time to client’s request time for the frame fetch. Where as in Server-Push, the server serves the client implicitly in response to the request of the client. The server is responsible for streaming the data in rounds and keeps track of the status of each stream. It ensures all the frames are streamed on time within the round time.

The challenges faced in designing the multimedia streaming servers are that the multimedia data requires totally different techniques for their organization and management when compared with the numeric and the text. The most critical of these is the continuity requirement. It becomes the responsibility of the multimedia streaming server to ensure that recording and retrieval of media streams with respect to disks proceed at real-time rates [8]. Designing a dedicated multimedia server that optimizes the client request service time is a matter of challenge.

Multimedia servers are connected to the clients via ATM (Asynchronous Transfer Mode) networks. Here the total bandwidth from the storage devices of the server via the network to the clients is fixed. A multimedia storage server can support only a limited number of clients simultaneously. The major concern for a server is to provide service with a good quality to a numerous groups of clients keeping the server and its network resources within feasible limits. An admission control algorithm determines the acceptance or rejection of a new request. It checks if the available bandwidth is sufficient for the total bandwidth required by the streams currently being serviced and the bandwidth requirement of the new request [15]. Our approach is exclusively for the clients who read from the disk and not write In this paper we have proposed dynamic buffer allocation strategy based on popularity-aware interval caching for prefix, which stores only the prefix of most popular multimedia objects and it batches the requests without the QoS violations of the request and does multicast transmission for all the clients. This reduces the overhead of the hard disk; it increases the number of concurrent users and utilizes the network and disk bandwidth efficiently.

The organization of the paper is as follows.
Section 2 is a review on some of the existing works on different Multicast, Popularity and Prefix aware interval caching scheme. Section 3 describes a Distributed VoD architecture Section 4 presents the importance of dynamic buffer allocation as compared with static buffer allocation. Section 5 presents a Multicast VoD Architecture. Section 6 presents the proposed algorithm. Section 7 the simulation results and discussion are presented. The conclusion and the future work are presented in section 8.

## 2. Related Work

To improve the multimedia streaming services, many studies on the caching of the multimedia streaming objects have recently been studied.
Among this Dan and Sitaram proposed a caching scheme for VoD servers named interval caching (IC), which exploits the temporal locality of accessing the same multimedia object consecutively [3]. The interval caching scheme consist of all consecutive requests pairs by the increasing order of memory space requirement, and then allocates memory space to as many of the consecutive pairs as possible. When an interval is cached, the following stream could read the data without any disk access since it will be directly served from the buffer cache.
Nachum et al. proposed a Real-time multicast Communication admission control procedure [1,12], which considers requests of multiple streams from multiple destinations and resolve contention when users requests exceeds the available network resources. However, the interval caching based on popularity and to retain the prefix, the initial portion of the video as opposed to later is not considered.
Ohhoon Kwon [2] has proposed a popularity and Prefix Aware Interval caching for multimedia streaming servers, but they have not extended their work on the admission control and increasing the number of concurrent users by multicast transmission.
Mcache [13] proposes a technique to remove the initial payouts delays of clients in multicast-based video streaming. While requests are batched together for a multicast, clients can receive the prefix of a requested movie clip from caches located in their own regions. But we have extended the work by adding a dynamic buffer allocation scheme.

## 3. Distributed VoD System

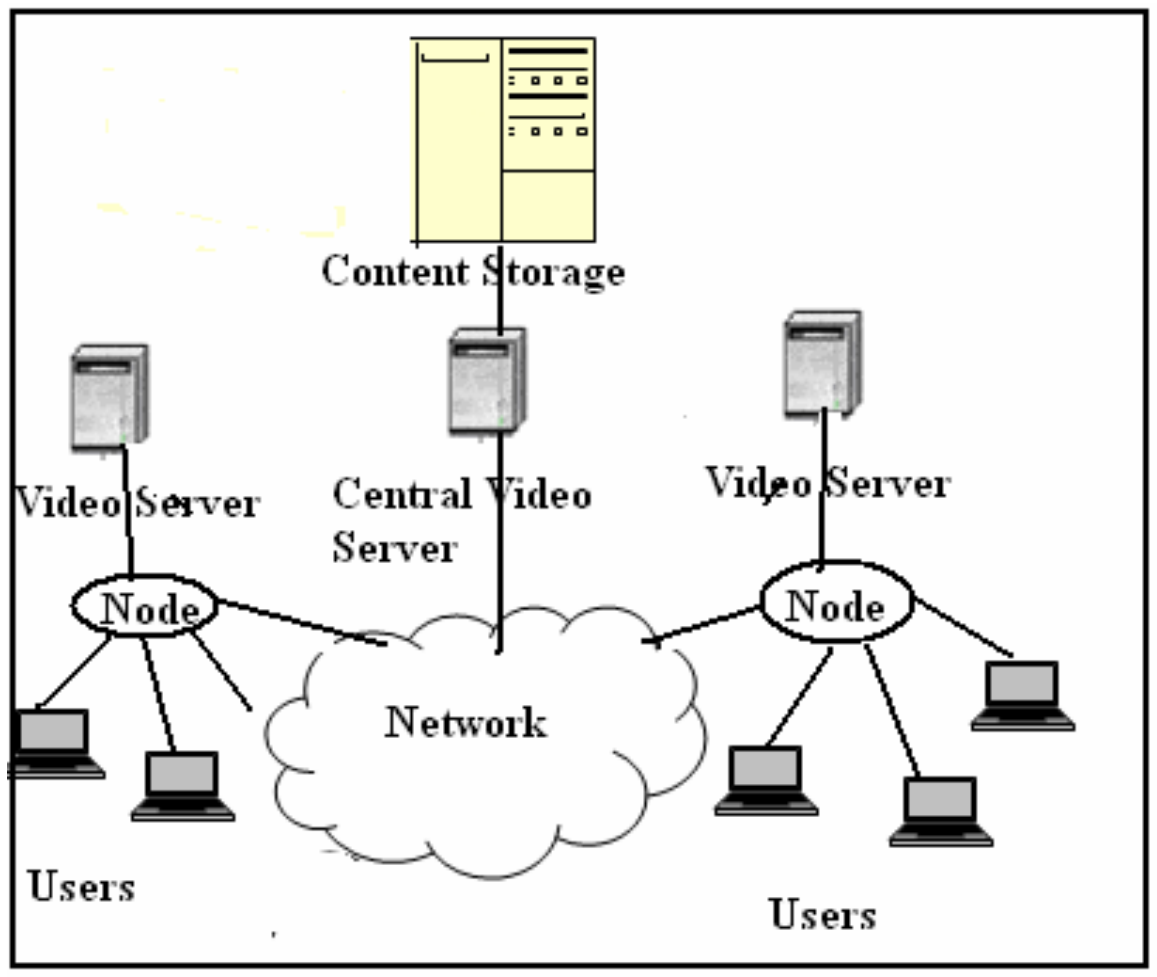

Figure 1: Distributed VoD System

In a distributed configuration there is one central server that stores all the content with smaller servers located near the network edges that are used to store high demand content (Figure 1). When a client requests that a particular video be played, the video server responsible for the request reserves sufficient processing capacity and network bandwidth for the video stream to guarantee continuous playback of the video. Transmission of video streams requires high network bandwidth and this can be very expensive, especially in the core section of the network.

## 4. Dynamic buffer allocation

Video data is provided to users by the VoD system. There are two important characteristics of video data. First, the amount of video data is voluminous. Second, video data must be continuously provided to the user.
The former requires that VoD systems use buffers for managing data by block units because systems cannot store the entire video data in memory. The latter mandates buffer management of VoD systems to retrieve new data blocks into the buffer before a user request uses up the data in the buffer. Minimizing memory requirement and initial latency is an important factor in buffer management of VoD [9].
Initial latency is the duration between the arrival of a user request and the arrival of the requested video data in the server's main memory. By dynamically allocating memory blocks based on popularity a number of concurrent user requests can be serviced with the same amount of memory.
To minimize memory requirement and initial latency several buffer scheduling methods has been proposed [10,11]. The buffer scheduling method determines the order of filling data buffers allocated to user requests.

These methods use static buffer allocation to allocate buffers to user requests. The static buffer allocation scheme determines the minimum buffer size based on the assumption that the system is in the fully loaded state, i.e.,the system services the maximum number of user requests that can be supported. The system consistently allocates this buffer size to all user requests regardless of the system's load. VoD systems must allocate larger buffers to user requests as the number of user requests in service increases.

The static scheme increases memory requirements and initial latency of systems [9,10], since it has a disadvantage that it uses memory inefficiently by allocating a large buffer than necessary, when the system is not in the fully loaded state.

## 5. The Architecture of Multicast VoD System

In video delivery system the popularity and access pattern plays an important role. As different videos are requested at different rates and at different times, videos are usually divided into the most popular and less popular requests. The top 10-20 videos are known to constitute 60-80% of the total demand.

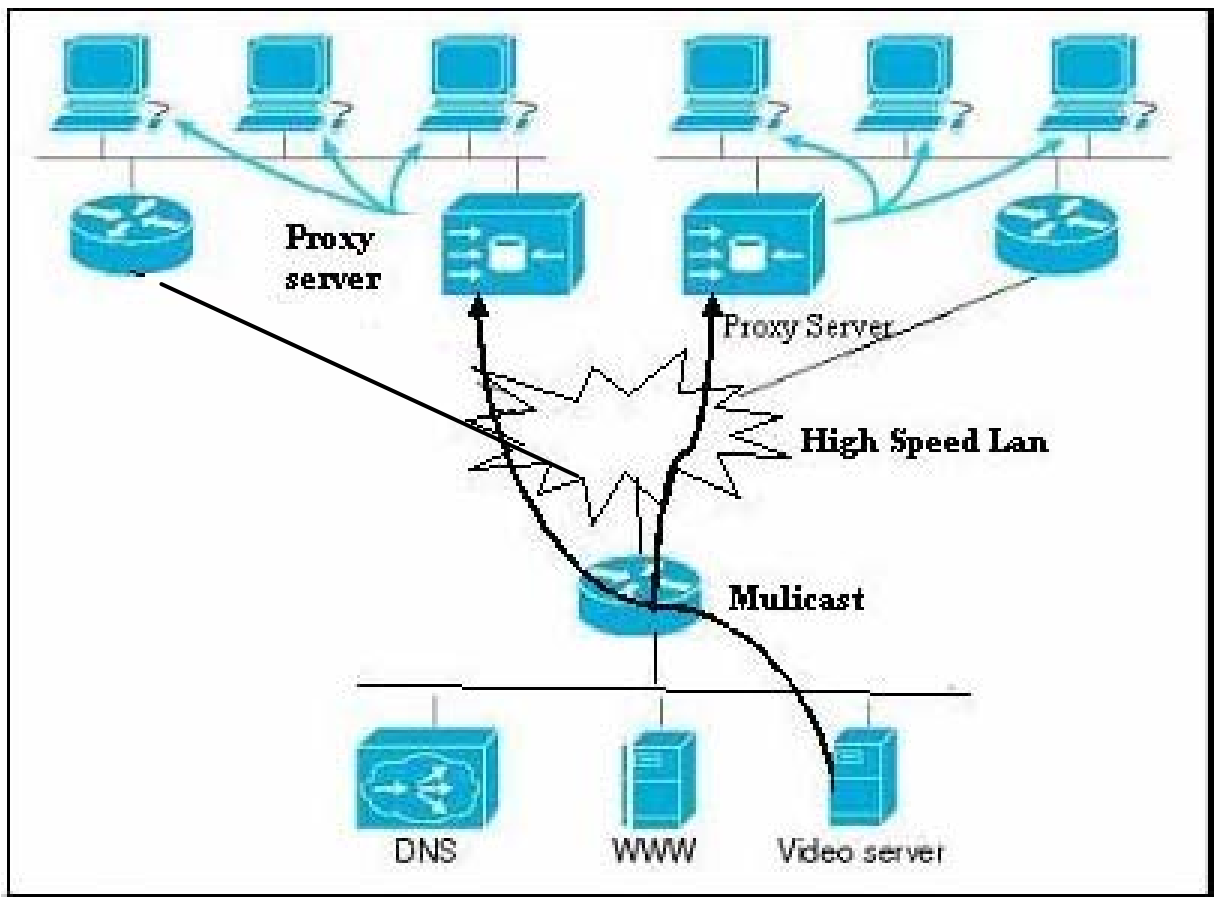

Figure 2. A Multicast VoD System

So, it is crucial to improve the service efficiency of most popular videos. Thus, requests by multiple clients for the same video arriving within a short time interval can be batched together in the cache and serviced using a single stream. This is referred to as batching.

The Multicast facility of modern communication networks offers an efficient means of one-to-many data transmission as shown in figure 2. The basic idea is to avoid transmitting the same packet more than once. Since all the local servers are connected to the content servers, as the popularity of any video increases the number requests arriving from the local server to the content server also increases [5]. The same copy of prefix of the video can be multicasted to all the local servers and sometimes directly to the clients by batching the entire request together with a small start up delay due to batching. This significantly improves the VoD performance, because it reduces the required network bandwidth.

All hosts that join a multicast group will receive the multicast traffic assuming that multicasting is supported in general. The idea behind controlled multicast framework is to control the multicast receivers and sources for certain multicast groups. This control can be done by the internet service operator or it can be done based on application specific needs e.g. by the multicast source [14].

In spite of these advantages, Multicast VoD has the following challenges

It is difficult to maintain the VCR (Video Cassette Recording) like support. Batching makes the clients arriving at different times to share a multicast stream, which may incur long service latency. The routers should support the multicast of VoD.

Some of these challenges are overcome in this paper. Usually VCR-like support is not expected from the proxy servers. The service latency problem is overcome by starting the service to the whole batch within the deadline constraints of the first request as shown in figure 3.

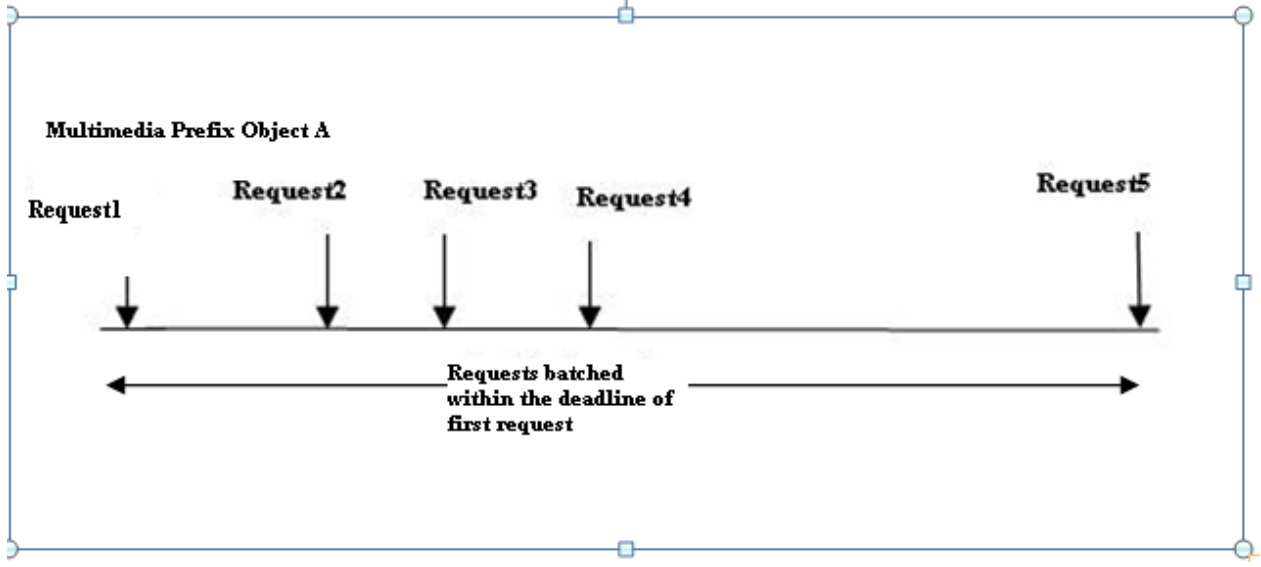

Figure 3. Batching of requests

This paper proposes a dynamic buffer allocation scheme, a novel approach for the buffer allocation that dynamically allocates the minimum buffer size in the completely present state. The inherent difficulty in allocating the buffer in the dynamic buffer allocation scheme is that the size of the buffer currently being allocated is dependent on the number of and size of the buffers to be allocated in the future, which are yet to be determined. We provide a solution to this problem using the popularity based algorithm to be described in Section VI. The advantages of this scheme are as follows: First, this scheme removes the static buffer allocation scheme’s problem of allocating unnecessarily large buffers. Second, by allocating the minimum buffer size for least popular video’s prefix, our scheme significantly improves the average initial latency and the average number of concurrent user requests that can be supported. Third, this scheme is independent of buffer scheduling methods utilizing all the buffers allocated. The buffer scheduling method determines the order in which the server fills the buffers with the data.

## 6. Multicast and Interval Caching

Whenever the bandwidth bottleneck limits the number of clients a multimedia streaming server can simultaneously support, a multicast delivery with interval caching [3, 4] becomes particularly attractive for multimedia applications. Interval caching exploits the high skewness in video access patterns by attempting to pair each playback request with an immediately preceding request for the same video that is currently being serviced from the cache. Specifically, interval caching reuses the data brought by a stream in servicing a closely following stream. The two streams are called the following stream and the preceding stream, respectively.

Only the initial portion of a media object, called the prefix, is streamed from the server's cache. Batching is an approach used to exploit the memory bandwidth and to save disk bandwidth in media servers by defining temporal cycles called batching windows. All requests that arrive within such a cycle are collected and at the end of the cycle, all the requests to the same video are serviced from the same media object saved in the cache. Upon receiving a continuous request from the client, the server collects all the requests within a cycle and immediately delivers the prefix to the client at the end of the cycle. The basic approach is the creation of a multicast group for the delivery of a video stream to the requesting end-user. If another user requests the same video shortly after the start of this transmission, the request is added to the same multicast group. This makes the cache sharable between as many clients as possible. Half portion of the cache is used to store only the prefix of the most popular videos. Naturally, an increase in cache hit percentage would have a direct bearing on the number of clients that can be admitted and served without missing deadlines. Our approach thus balances the time duration of a batching window and deadline of each requested video.

Moreover, the number of initial segments cached is dynamically determined by the popularity of an object. It is assumed that the latency between a client device and the local server is negligibly small, but the latency between the local server and the content server is relatively large and cannot be ignored. Here, we assume that we are dealing with only clients that are requesting the data to retrieve from the disk and not storing to the disk.

The retrieval time directly from the server and the start up time can be is remarkably reduced when the prefix is directly streamed from cache. But a small latency time is incurred while batching all the requests.

As an example, recent studies found that nearly 90 percent of media playbacks are terminated prematurely by clients after watching the initial portion of the video. Hence fetching the remaining portion i.e the suffix, from the server's disk to the prefetch cache relays on the client's interest to continue to watch or not.

## 7. Proposed Algorithm

We have assumed a multicast transmission with interval caching. Whenever a client request for a video, the requested video's prefix may be or may not be present in the local server.

If the requested video's prefix is present in the local server, then the real time transmission of the video starts immediately with the video content being streamed to the client from the local server. If the requested video is being streamed from the central multimedia server to the local server, then also the real time transmission of the video starts immediately with video content being streamed to the client from the local server. If the requested video is not present in the proxy server the following dynamic buffer allocation occurs:

All the videos have an associated number (popularity) which gives the popularity of the video based on the number of hits. i.e. the popularity of a video is directly proportional to the number of hits for that video. Initially when all the blocks are free, the required number of blocks based on the popularity for the video is allocated. If the required number of blocks is not available, we find the prefix present in the cache and is currently not being streamed (completely offline) and is allocated more than the minimum number of blocks. If the popularity of the requested video's prefix is more than the popularity of lowest popular video having more than the minimum number of blocks, then the blocks except minimum number of blocks are deallocated from the lowest popular completely offline video and these deallocated blocks are allocated to the requested video. The same procedure will be repeated considering the next least popular prefix of video until the requested video gets required number of blocks based on the popularity.

If we cannot find the lowest popular video which is completely offline in the local server, then we find the lowest popular video, which is present in the local server and is currently being streamed and is allocated more than minimum number of blocks. If the popularity of the requested video is more than the popularity of lowest popular video, which is currently being streamed having more than minimum number of blocks, then the blocks except minimum number of blocks are deallocated from the lowest popular completely and then these deal located blocks are allocated to the requested video. The same procedure will be repeated considering the next least popular video until the requested video gets required number of blocks based on the popularity.

If the requested video gets at least the minimum number of blocks then assign the allocated number of blocks. If we cannot find the lowest popular video's prefix which is currently being streamed having more than minimum number of blocks in the cache and the requested video does not get even the minimum number of blocks then the request is rejected.

We assume the minimum size of the cache allocation unit as c, and all allocations are in multiples of this unit. It can be one bit or minute's worth of data, etc. We express the size of video and the prefix cache size as a multiple of a unit c. Video has playback bandwidth $b_i$ bps, $L_i$ length seconds, and size $n_i$ units,

$n_i c= b_i L_i$. We assume that the cache can store storage vector V =($v_1$,$v_2$,$v_3$….$v_n$) specifies that a prefix of length $v_i$ seconds for each video is cached at the proxy. Note that the prefix videos cached at the proxy cannot exceed the storage capacity of the proxy.

## Algorithm

```
When a request Run for prefix arrives at time t
If (No. of cache blocks free > the no. of cache blocks required by Run based on the popularity)
 {
   Assign the required no. of cache blocks
   to Rn.
 }
else
 {
   Assign the available no. of cache blocks
   to Rn.
   Identify the set of request {Ri} for a prefix interval caching with multicast group which are completely offline having more than minblks
     If (found) (allocate the cache blocks to Rn)
      While(cache blocks not completely
      allocated to Rn based on popularity
      && still there are some videos in interval caching with multicast group popularity<Rn)
        {
          consider the least popular video
          Ri of a multicast group
          if(popularity of Rn>popularity of
             Ri)
          {
            free except minblks from the
            request Ri
            add these blocks to Rn
          }
        }
      if(Rn does not get the required no. of
      cache blocks based on popularity)
        {
        Identify the set of request{Ri} in interval caching in a multicast group which are being streamed and having more than Min no. of blocks
        If (found)
         {
           While(cache blocks not
           completely allocated to Rn &&
           still there are some videos
              with popularity<Rn)
            {
              consider the least popular
              video Ri
              If(popularity of
                 Rn>popularity of Ri)
                {
                  free except minblks from
                  the request Ri
                  add these blocks to Rn
                }
            }
         }
        else
          if (Rn gets at least min no. of
              blocks )
            {
              assign the allocated no. of
              blocks to Rn .
            }
          else
            {
              Reject the request.
            }

        }
 }
```

## 7. Results and Discussion

In any distributed VoD system, each node can serve a large number of homes (nearly 20,000). A hub may also exist beyond a node which is connected via a fibre which in turn can serve nearly 750 homes. An HFC network typically has a total spectrum of 750MHz. This in turn is divided into individual analogue channel of 8MHz each. These channels are used for services such as high speed internet and pay TV. If a VoD service is deployed on an HFC (Hybrid Fibre Co-axial) network a fixed number of analogue channels are assigned for use. Using QAM-64(Quadrature Amplitude Modulation) one 8MHz can transmit approximately 38Mbps. If the video content is encoded at 3.5Mbps (most commercial VoD systems use video content at this rate), then one analogue QAM channel can support approximately 10 digital video streams concurrently. Therefore if 4 QAM channels are assigned for VoD, then a maximum of 40 requests can be served at each node at any given time during peak period. Hence it is required to utilize a maximum limit of concurrent request capacity of any node to increase its efficiency.

This is a model taken for and appropriate values can be selected for any system depending on current technology and capital investment.

The results presented below are an average of several simulations conducted on the model. The values considered for simulation are as follows: Size of one cache block = 1MB, Total cache blocks considered in the proxy server = 1500, Size of the videos = U (300MB, 500MB), Minimum numbers blocks allocated to a video = 30MB, each simulation is carried out for 1500 seconds. Our simulation model consists of a central server with 100 complete movies stored in it. A cache has the space to hold 10 movies. 50% of the cache is used to store only the prefix and remaining holds the later parts of the movie being streamed the central content server is connected to the proxy servers with high speed networks (ATM).Total start up latency of the hard disk is approximately 6 ms. Mean number of blocks per video is 200MB, mean inter arrival time is 60 s. Maximum hard disk bandwidth 10Mbytes/sec.

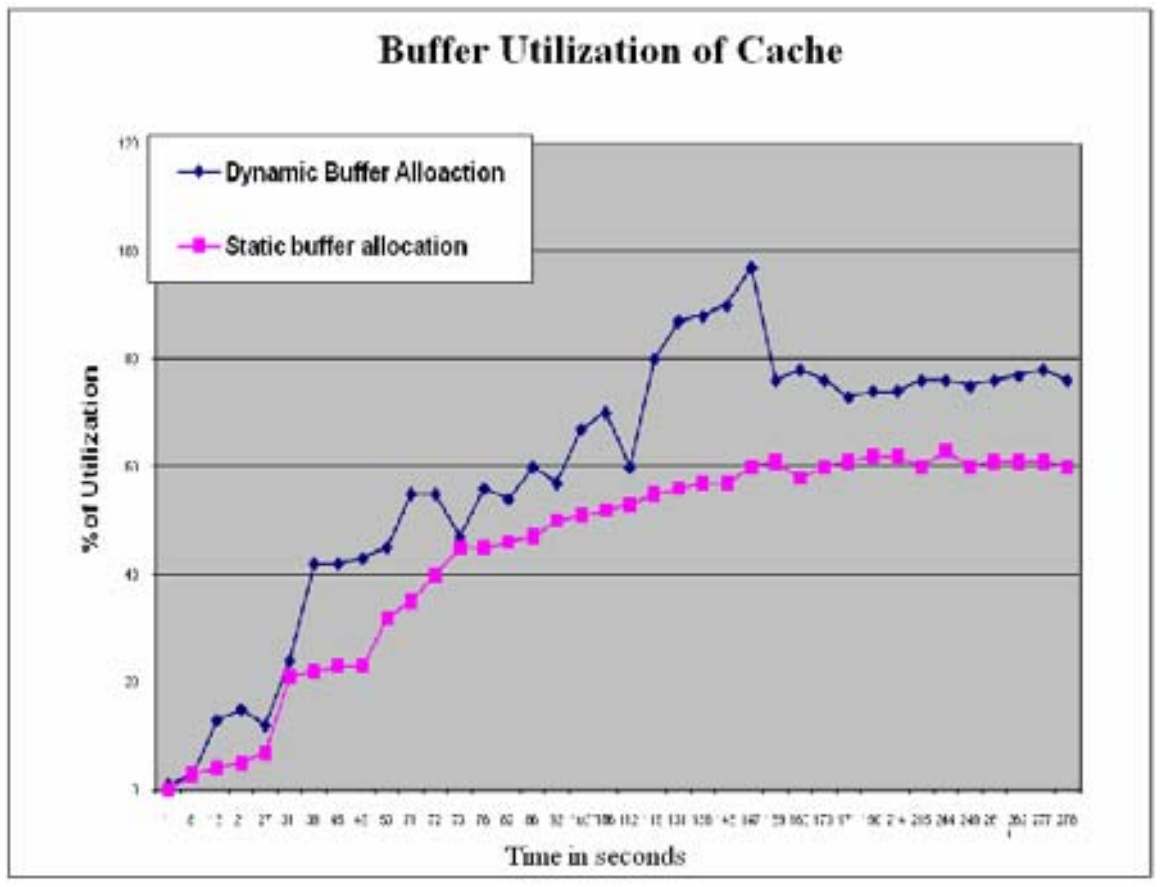

Figure 4 Buffer utilization

Figure 4 describes the how storing the prefix of the stream can improve the delivery of continuous media and increases the buffer utilization. Buffer utilization can be 100% when most of the popular videos get at most minimum blocks of cache.

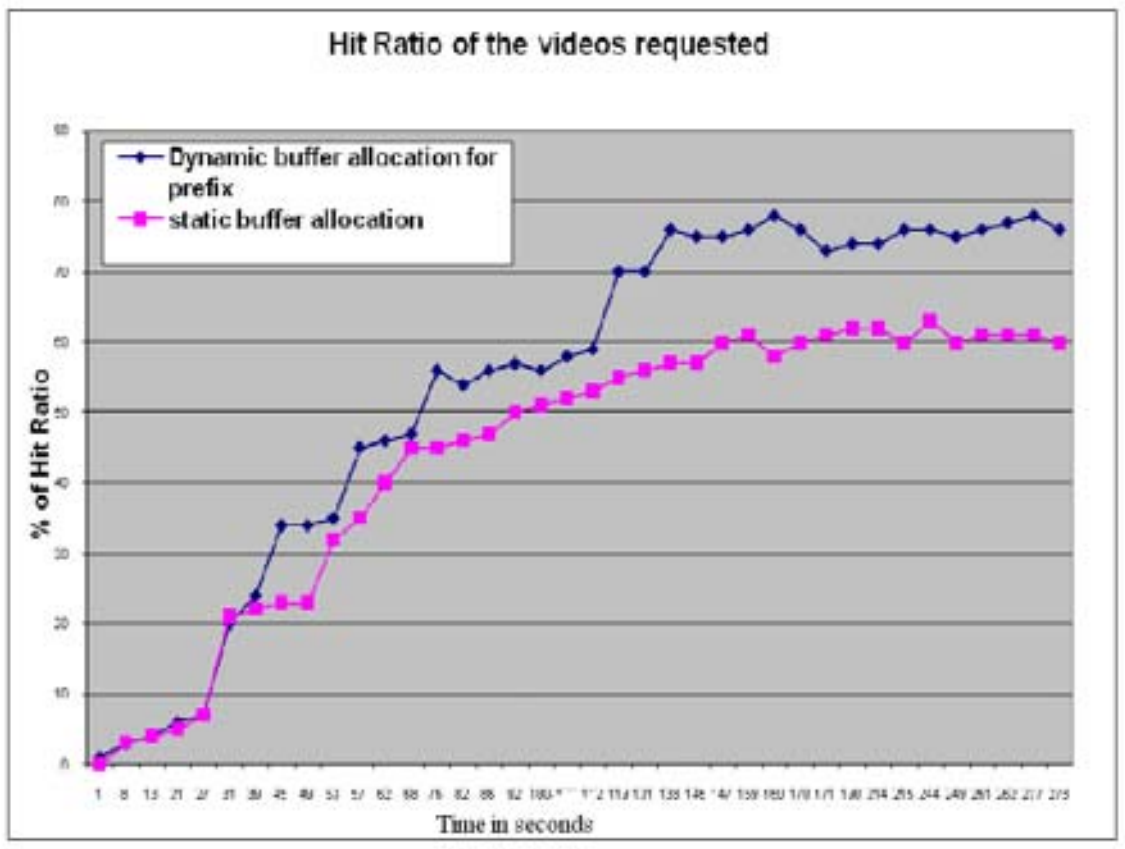

Figure 5. Hit Ratio of the Videos

Figure 5 shows that hit ratio of videos are increased in dynamic buffer allocation since more number of most popular video is stored in the cache .

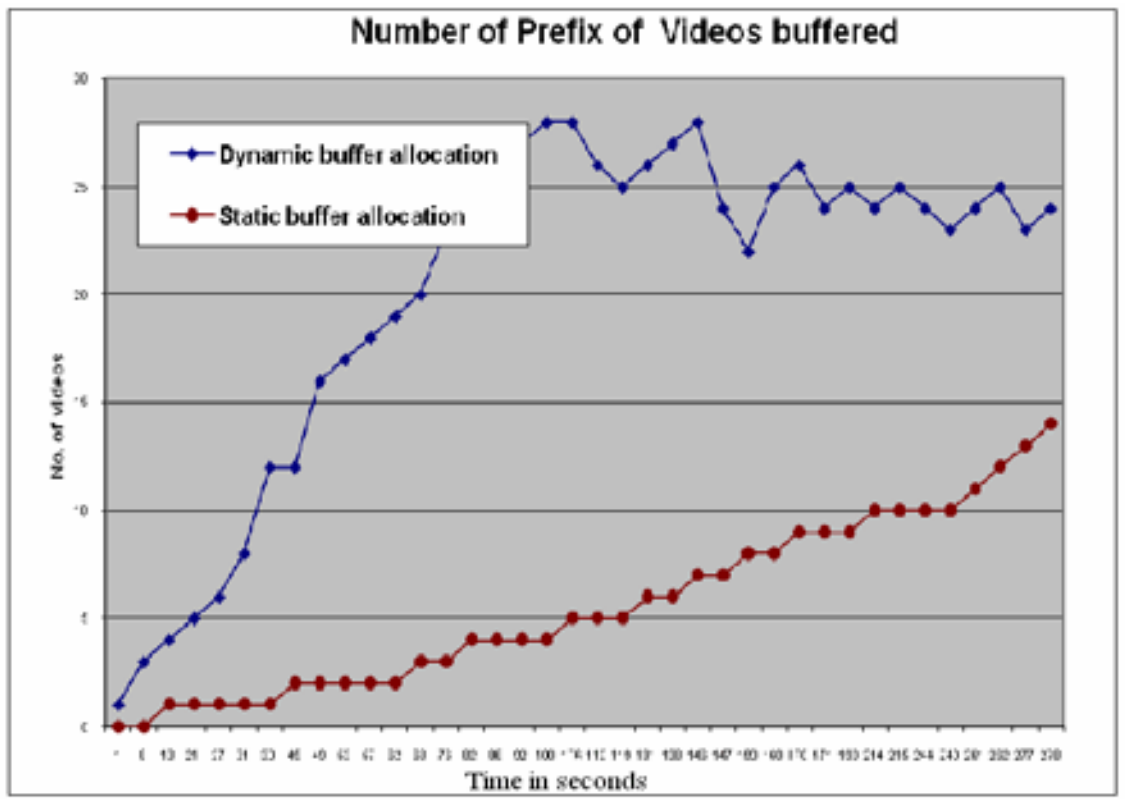

Figure 6. Total Videos buffered

Figure 6 show that existing static buffer allocation scheme determines the buffer size assuming the fully loaded system state. Thus, the static scheme allocates an unnecessarily large buffer when the system is not in the fully loaded state. In contrast, the dynamic buffer allocation scheme allocates the minimum buffer size in a partially loaded state, as well as in the fully loaded state. Hence increases the number of prefix that can be stored as compared with static buffer allocation.

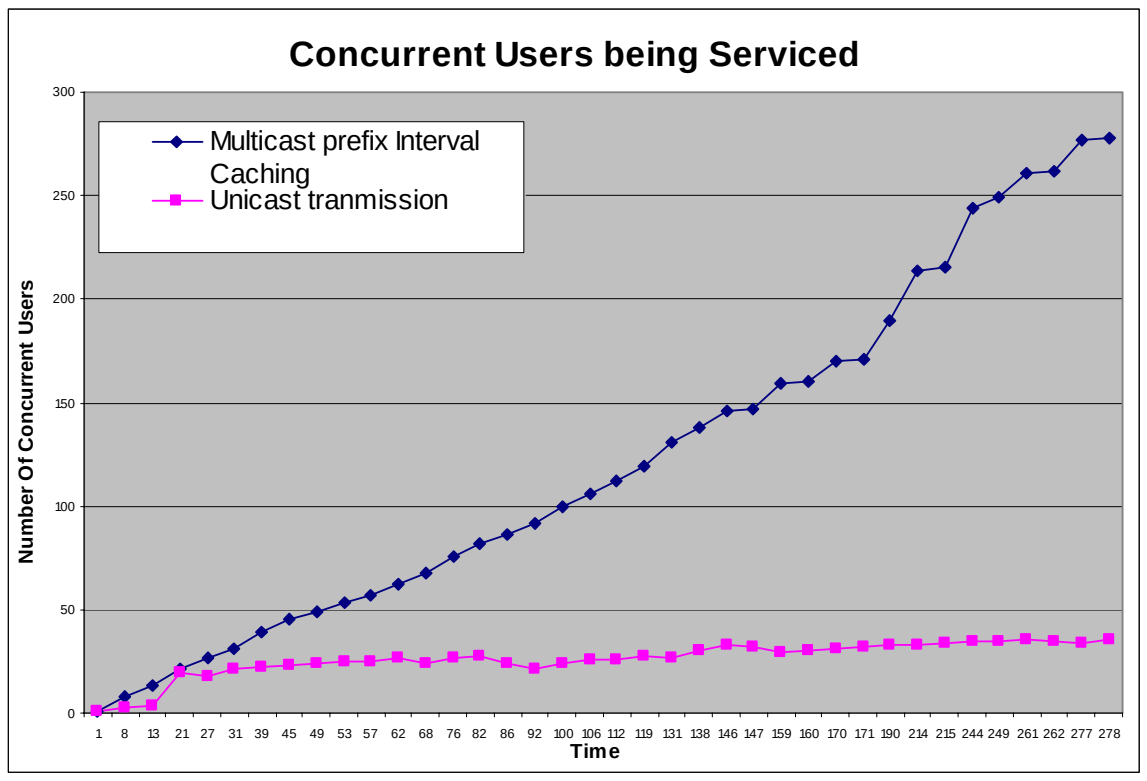

Figure 7: Total number of concurrent users

Figure 7 shows; the maximum number of concurrent users in unicast transmission is limited to 40. Multicast enables hundreds users to play a single stream, provided the streaming server work on networks with Multicast-enabled routers.

## 8. Conclusion and Future work

In this paper we have shown that by storing only the prefix of most popular video in the cache and batching all the requests with the time interval of deadline of first request and then multicasting the video to all the requested clients we provide admission for more clients. Our work is compared with the static buffer allocation for the prefix. With dynamic allocation of buffer based on popularity fix we not only get more number of videos being streamed but also reduction in the rejection ratio. The request to service time is also reduced. As the number of most popular videos stored in the cache increases, the ratio also increases. The future work can be carried out for a further improved replacement technique to increase the hit ratio using bandwidth to space ratio.

**P Jayarekha** holds M.Tech (VTU Belgaum) in computer science securing second rank. She has one and a half decades experience in teaching field. She has published many papers. Currently she is working as a teaching faculty in the department of Information science and engineering at BMS College Of Engineering, Bangalore, India.

**T.R. Gopalakrishnan Nair** holds M.Tech. (IISc, Bangalore) and Ph.D. degree in Computer Science. He has 3 decades experience in Computer Science and Engineering through research, industry and education. He has published several papers and holds patents in multi domains. He won the PARAM Award for technology innovation. Currently he is the Director of Research and Industry in Dayananda Sagar Institutions, Bangalore, India.